\documentclass[11pt, onecolumn,draftclsnofoot]{article}
\usepackage[super,sort&compress,comma]{natbib}
\usepackage{multirow}
\usepackage{changepage}
\usepackage{color}

\usepackage{mathrsfs}
\usepackage{pifont}
\usepackage{bbding}
\usepackage{tipa}
\usepackage{cite}
\usepackage{epsfig}
\usepackage{graphicx}
\usepackage{url}
\usepackage{amsfonts}
\usepackage{multicol}
\usepackage{float}
\usepackage{times}
\usepackage{psfrag}
\usepackage{subfigure}
\usepackage{stfloats}
\usepackage{footnote}
\usepackage{array}
\usepackage{amsmath,epsfig}
\usepackage{stmaryrd}
\usepackage{amssymb}
\usepackage{epic}
\usepackage{curves}
\usepackage{bm}
\usepackage{balance}
\usepackage{caption}
\usepackage{subfloat}
\usepackage{subfig}
\usepackage{graphicx}
\usepackage{subfigure}
\usepackage{algorithm}
\usepackage{algpseudocode}
\usepackage{amsmath}
\begin{document}

\title{Error-Control-Coding Assisted Imaging}


\author{Xiaopeng~Wang$^{1}$,
\thanks{e-mail: xiaopeng.wang@sydney.edu.au. $^1$Center of Excellence in Telecommunications, The School of Electrical and Information Engineering, The University of Sydney, NSW, 2006, Australia. $^2$Key Laboratory for Quantum Optics and Center for Cold Atom Physics of CAS, Shanghai Institute of Optics and Fine Mechanics, Chinese Academy of Sciences, Shanghai, 201800, China.}
\and Zunwang~Bo$^{2}$,
\and Zihuai~Lin$^{1}$,
\and Wenlin~Gong$^{2}$,
\and Branka~Vucetic$^{1}$,
\and Shensheng~Han$^{2}$
}
\date{}
\maketitle

\begin{abstract}
Error-control-coding (ECC) techniques are widely used in modern digital communication systems to minimize the effect of noisy channels on the quality of received signals. Motivated by the fact that both communication and imaging can be considered as an information transfer process, in this paper we demonstrate that ECC could yield significant improvements in the image quality by reducing the effect of noise in the physical process of image acquisition. In the demonstrated approach, the object is encoded by ECC structured light fields generated by a digital-micromirror-device (DMD) while its image is obtained by decoding the received signal collected by a bucket-detector (BD). By applying a Luby Transform (LT) code as an example, our proof-of-concept experiment validates that the object image can be effectively reconstructed while errors induced by noisy reception can be significantly reduced. The demonstrated approach informs a new imaging technique: ECC assisted imaging, which can be scaled into applications such as remote sensing, spectroscopy and biomedical imaging.

\end{abstract}

Error-control-coding (ECC) technique has its origin in the pioneering work of Shannon's published in 1948. In his well-known paper \emph{A Mathematical Theory of Communication}, Shannon proved that by properly encoding the information, one can achieve any arbitrarily low error probability, as long as the data transfer rate does not exceed the channel capacity\citep{shannon2001mathematical}. Then in the following decades, a variety of ECC codes are invented to combat errors caused by noisy channels, providing efficient and reliable digital data transmission in commercial, governmental and military spheres\citep{hamming1950error,reed1960polynomial,bose1960class,gallager1962low,berrou1993near,luby2002lt,arikan2009channel,shirvanimoghaddam2013near}. Though there are many different classes of ECC, their underlying principle is the same, i.e. adding structured redundancy at the transmitter. By following deliberately designed encoding rules, redundancy is created and added into the transmitted encoded symbols. Thus even if some of the received symbols are corrupted by noise, the decoder can use the redundancy information to recover the original message with a high probability.

\begin{figure}
  \centering
  \includegraphics[width=5in]{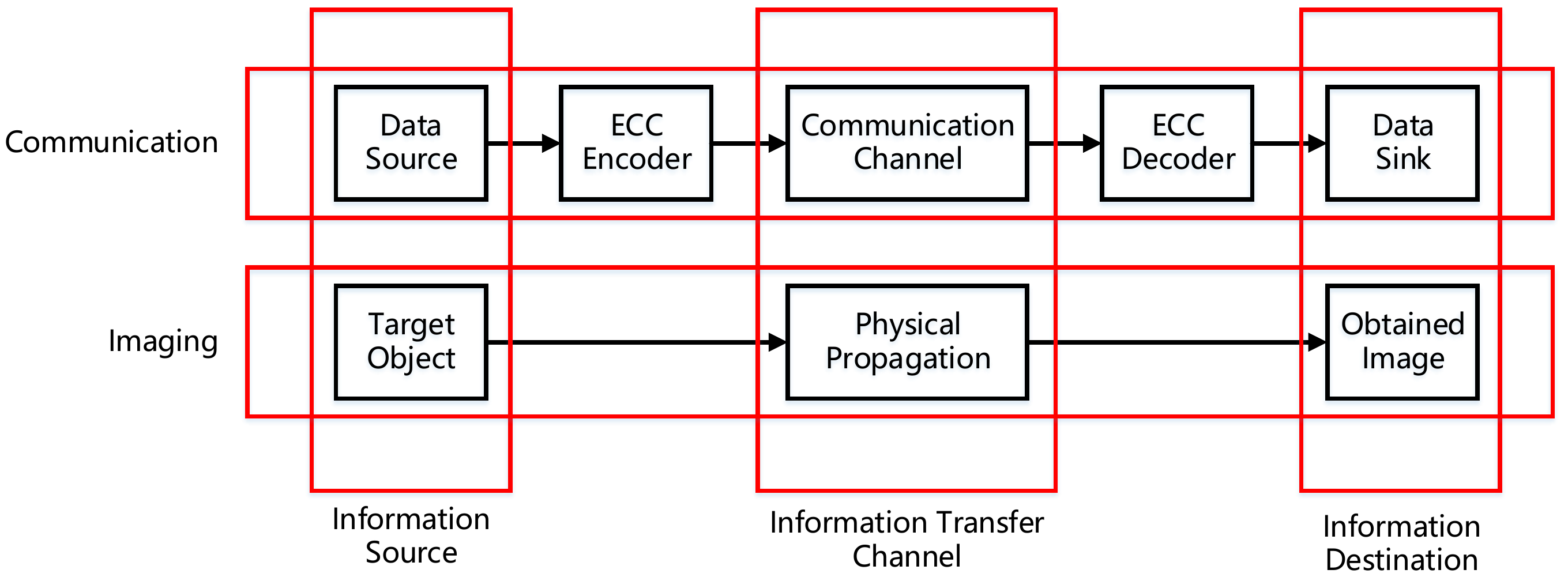}
  \caption{\textbf{A block diagram of a digital communication systems and an imaging systems.} Both of them contains an information source, an information transfer channel and an information destination. In communication systems information is encoded by ECC before transmission while there is no similar procedure in imaging systems.}\label{ImagingandCommunications}
\end{figure}

Imaging is a physical procedure which acquires the representation or reproduction of an object such as its shape, location, structure, etc. Radiations and fields such as electromagnetic waves, particles and quasistatic fields can be used to form an image\citep{barrett2013foundations}. However, no matter what physical means an imaging system applies, it is an information transfer process which can be represented by the same basic model as a communication system. As shown in Fig. \ref{ImagingandCommunications}, communication and imaging systems both contain an information source (data source in a communication system or target object in an imaging system), an information transfer channel (communication channel in a communication system or physical propagation in an imaging system) and an information destination (data sink in a communication system or obtained image in an imaging system). One of the major differences between them lies in the information processing before transmission. In communication systems, information is encoded by an ECC and transmitted over the channel. When errors occur during the information transfer process, communication systems can achieve a high reliability with the assistance from ECC codes. On the other hand, the object information can also be corrupted by noise presented in an imaging system. However, in its current framework, there is no similar procedure so far. Thus by following the analogy with communication systems, we posed a logical question: Could ECC, developed for communication systems, be applied to imaging systems to protect object information against noise and achieve a more reliable imaging performance?

In this article, we provide an initial affirmative answer to this question. Under the assumption that the object information is represented by binary-valued transmissivity, a digital-micromirror-device (DMD) is used to encode the object by light fields generated according to ECC encoding rules. Then the transmit-through lights are collected by a bucket-detector (BD) and passed to the decoder. After converting analog BD outputs to the Galois field GF(2) where most ECC codes are designed to work, the object information is recovered by applying ECC decoding algorithms designed for communication systems. Our proof-of-concept experiment validates that the object information can be effectively protected against errors induced by reception noise. Under the same system settings, the superiority of the demonstrated approach is further confirmed by benchmarking with ghost imaging (GI)\citep{pittman1995optical,gatti2004ghost,cheng2004incoherent,jiying2010high}, which is chosen to represent conventional \emph{uncoded} imaging schemes.





\section*{Encoding Process In Imaging}\label{encoding}

\begin{figure}
  \centering
  \includegraphics[width=5in]{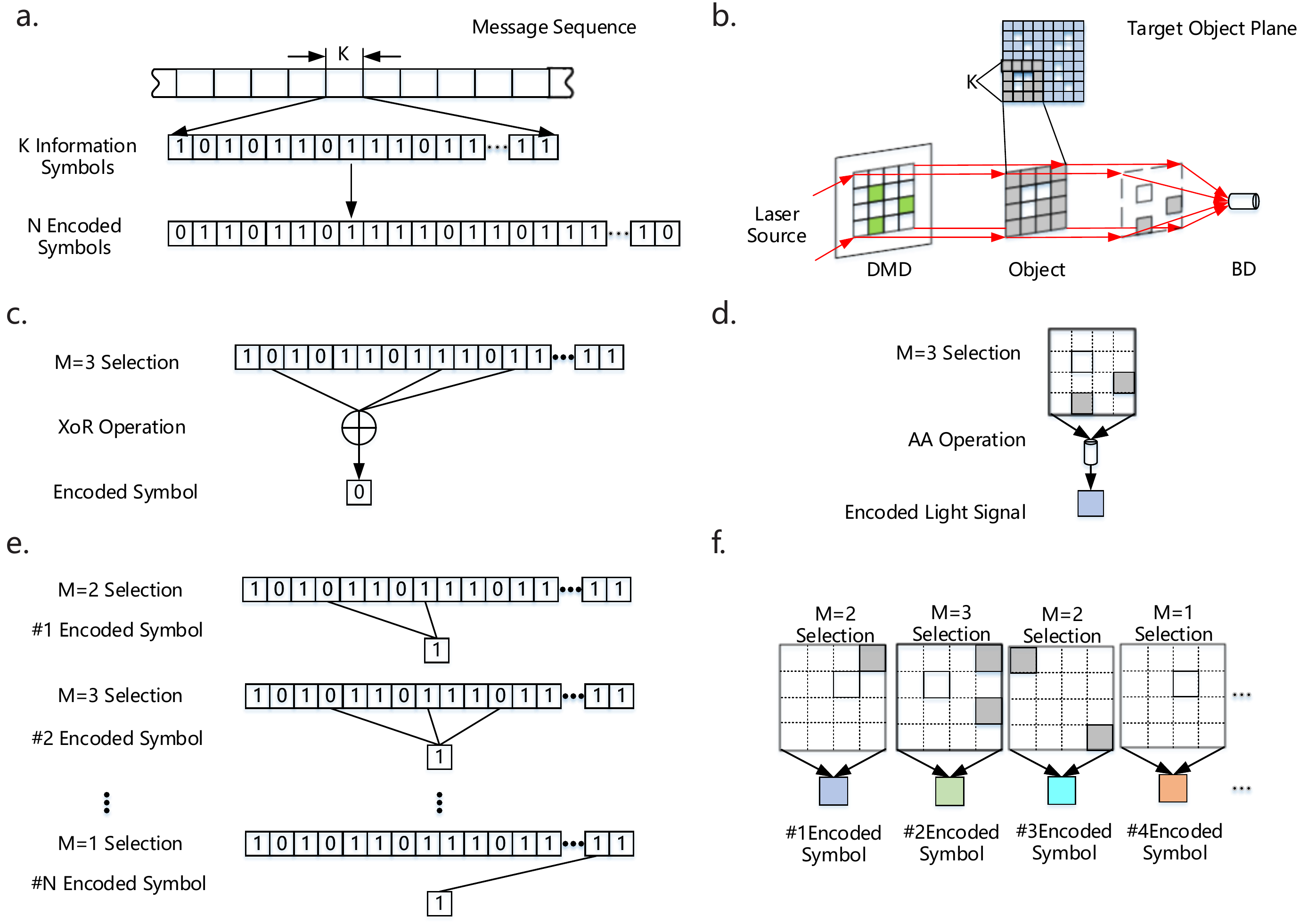}
  \caption{\textbf{Physical implementations of the encoding procedure.} \textbf{a}, The encoding operation in a communication system. \textbf{b}, The encoding operation in our design. \textbf{c}, The linear combination is realized by XoR operation in a communication system. \textbf{d}, The linear combination is realized by AA operations in our design and carried out by a BD. \textbf{e}, The generation of an encoded sequence in a communication system. \textbf{f}, The generation of an encoded imaging sequence in our design.}\label{EncodingProcess}
\end{figure}

In communication systems, the data source typically generates binary sequences, which are used as the input to the ECC encoder\citep{vucetic2003space,lin2004error}. The linear block code ECC encoding operation in a communication system starts with dividing a length $P$ message sequence into $Q$ blocks each consisting of length $K$ information symbols. As illustrated in Fig. \ref{EncodingProcess}a, each block with length $K$ is mapped into an encoded sequence of length $N$ by the ECC encoder, where $N>K$ for fixed rate codes\citep{gallager1962low} and $N=K$ for rateless codes\citep{luby2002lt}. This mapping is carried out by generating linear combinations of these $K$ binary information symbols over GF(2).

Though the target objects in imaging systems are physical entities represented by analog parameters, such as reflectance, refraction and transmissivity\citep{barrett2013foundations}, in order to simplify the application of ECC encoding we assume that the target object is represented by a binary-valued image, i.e. black and white. We choose transmissivity as an example of the physical representation of the object and divide the target object imaging plane into individual pixels whose values are defined to be either 1 or 0. Value 1 corresponds to the infinitely large transmissivity while value 0 corresponds to the infinitely small one. In such a manner, the information source in an imaging system, namely the object information, is quantized by binary-valued transmissivity and discrete pixel locations.


In our design, the encoding operation in imaging is performed as shown in Fig. \ref{EncodingProcess}b. By following the analogy with a communication system, we divide the whole target object plane containing $P$ pixels into $Q$ squares, where each of the squares consists of $K$ pixels. Then $K$ pixels from the same square are selected to perform $N$ times imaging encoding operations and generate an encoded imaging sequence of length $N$, where $N\geq K$. In each imaging encoding operation, $M_i$ $(i=1,2,\ldots N)$ out of $K$ pixels are chosen according to a specific ECC encoding rule\citep{gallager1962low,berrou1993near,luby2002lt}. Then a DMD is used to generate a light field to simultaneously illuminate the selected $M_i$ pixels with the same intensity and duration. After the generated light field has been modulated by the object, the transmissivities of $M_i$ illuminated pixels are included in the spatial intensity distribution of the transmit-through light. While the linear combinations of information symbols in a communication encoding operation is obtained by XoR operations\citep{vucetic2003space,lin2004error} as shown in Fig. \ref{EncodingProcess}c, these operations cannot be directly realized in our scenario since the resulting transmit-through light intensity is an analog quantity. Thus in order to achieve a linear combination of the light intensity, the XoR operations are decoupled into arithmetic additions (AAs) and modulo-2 operations in our design, where AAs are carried out by a bucket-detector (BD), as illustrated in Fig. \ref{EncodingProcess}d. The resulting sum, consisting of the total intensity of the transmit-through light, is corrupted by noise, as for example additive-white-Gaussian-noise (AWGN) caused by electronic components in the BD.

As shown in Fig. \ref{EncodingProcess}e, the encoding operation in communication systems is repeated $N$ times with different combinations of involved information symbols\citep{vucetic2003space,lin2004error} in order to generate the encoded sequence of length $N$. Similarly in our design, the above imaging encoding operation is also repeated $N$ times where the selection of $M_i$ pixels is determined by the encoding rule, as illustrated in Fig. \ref{EncodingProcess}f. The obtained encoded imaging sequence of length $N$ is passed to the imaging ECC decoder, which will remap the noisy sequence into GF(2) and then perform a decoding operation similar to ECC decoding methods in communication systems.

\begin{figure}
  \centering
  \includegraphics[width=5in]{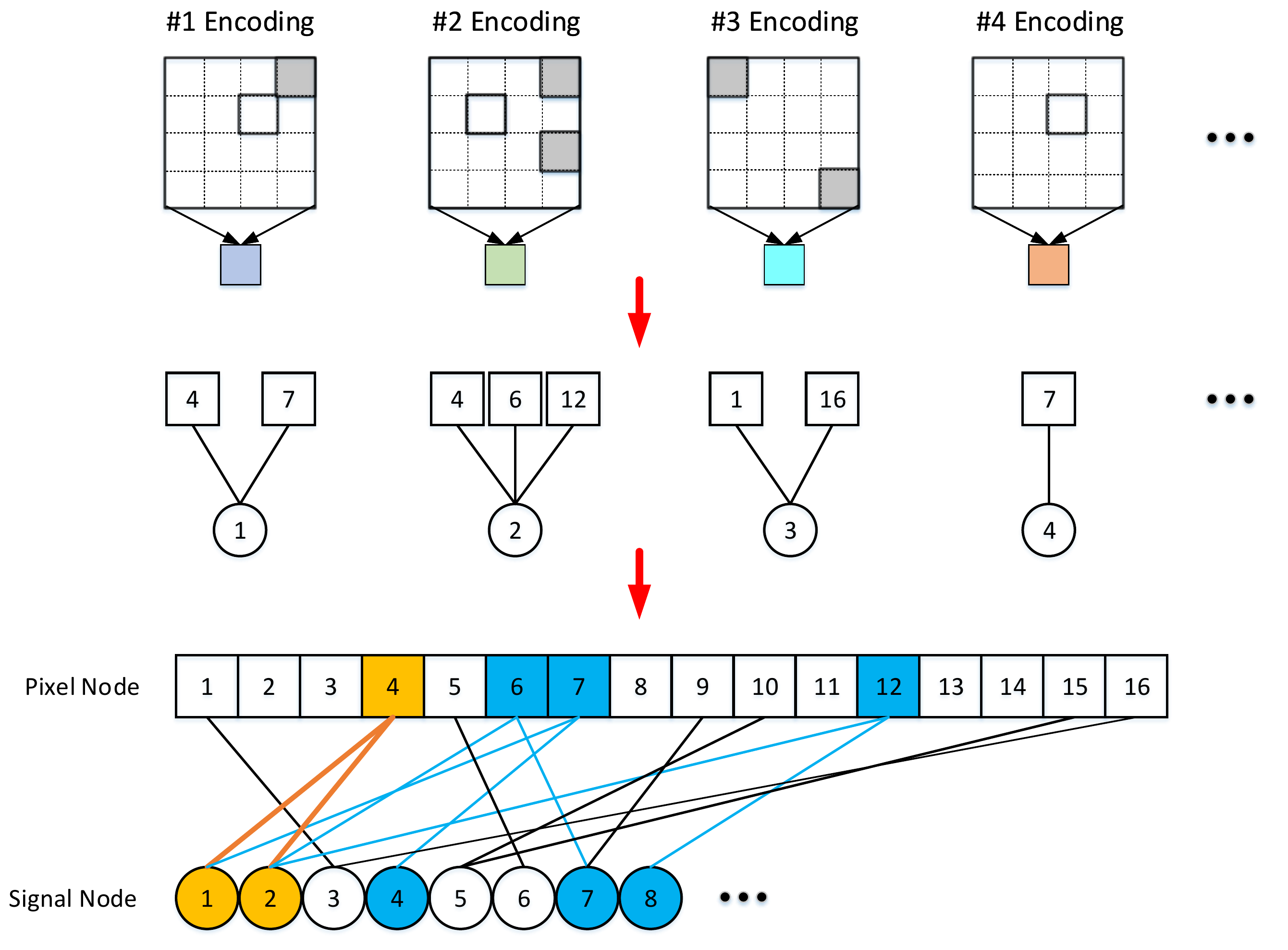}
  \caption{\textbf{The encoding procedure represented by a bipartite graph.} Circle-shaped nodes in the graph are representing received signal values and square-shaped ones are representing pixels. Numbers in squares are indicating the order of pixels while numbers in circles are representing the order of illuminations. Solid lines are expressing relationships between pixels and received signal values introduced by light fields generated by the DMD. Orange colored nodes are directly connected to each other via orange colored solid lines. Nodes colored in blue are providing extra messages to orange ones via blue colored solid lines.}\label{BP_explaination_Code}
\end{figure}

By using circle-shaped nodes to represent different encoded signal values and square-shaped nodes to represent different pixels, this physical encoding procedure can be further represented by a bipartite graph\citep{tanner1981recursive,asratian1998bipartite} in Fig. \ref{BP_explaination_Code}. Solid lines between circles and squares describe the relationships between received signal values and illuminated pixels established by light fields generated by the DMD. Since in one encoding procedure only the information from illuminated pixels is combined into the received signal value, the squares corresponding to the same illumination are connected to the same circle in the bipartite graph. This procedure is repeated $N$ times to generate a length $N$ encoded imaging sequence.

\section*{Decoding Process In Imaging}\label{decoding}

The imaging decoding consists of two main operations, namely the remapping of the received signal values into GF(2) and the decoding of object information.



\begin{figure}
  \centering
  \includegraphics[width=5in]{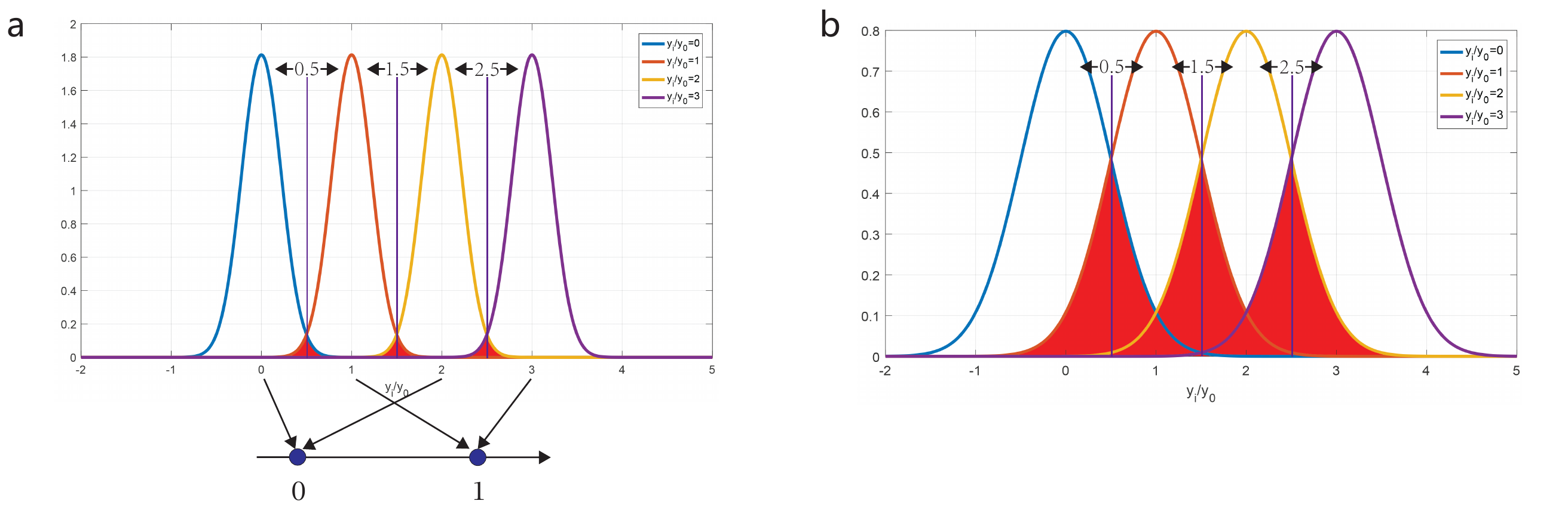}
  \caption{\textbf{The PDF of the received signal and its remapping.} \textbf{a}, The PDF of the received signal value when $N=3$ pixels are selected in the encoding procedure. A straightforward remapping scheme is to set thresholds at middle points between possible integers and round-off the received signal value to the closest one. Then after an additional modulo-2 operation, the received signal value can be remapped to GF(2). \textbf{b}, When the noise power becomes larger, intersections (red-colored areas) between different PDF curves will be enlarged. Thus more errors caused by decision ambiguity will be introduced into the remapping process.}\label{remapping}
\end{figure}


Assuming for pixels with value 1, the mean value of corresponding outputs from the BD is $\overline{y_0}$, ignoring propagation attenuations and only considering AWGN noise in the receiver\citep{barrett2013foundations}, the probability-density-function (PDF) for the received signal $y_i$ obtained from the $i$th imaging encoding procedure can be expressed as\citep{parzen1962estimation},
\begin{equation}\label{PDF}
  f(y_i|m\overline{y_0})=\frac{1}{\sqrt{2\pi \sigma^2}}\textmd{exp}(-\frac{(y_i-m\overline{y_0})^2}{2\sigma^2})
\end{equation} where $m=0,1,2,\ldots,M$ and $M$ is the total number of involved pixels in the $i$th encoding and $\sigma^2$ is the receiver noise variance. As the AA part of the XoR encoding operation is carried out by the BD, an intuitive remapping of the resulting sum is to find an operation that is equivalent to modulo-2. For example, when $M=3$ pixels are involved in the $i$th imaging encoding procedure, the PDF of the resulting sum $y_i$ output by the BD is as shown in Fig. \ref{remapping}a. To achieve the remapping, we can first normalize $y_i$ into $\widehat{y_i}=y_i/y_0$ and find an integer $m_i$ which is the closest to it. Under the assumption of the equal probability for each pixel value (either 0 or 1), this decision can be made by setting thresholds at middles points between possible integers (i.e. 0.5, 1.5 and 2.5). When the obtained $m_i$ is even, the corresponding value of $y_i$ on GF(2) is 0; if $m_i$ is odd, then $y_i$ corresponds to 1 on GF(2).

However, this \emph{hard-decisions} based remapping may bring extra errors, since red-colored intersections shown in Fig. \ref{remapping}a will cause decision ambiguity. This decision ambiguity will increase with noise, as shown in Fig. \ref{remapping}b. Consequently, \emph{soft-decisions} representing the likelihood for $y_i$ are employed in our design to enable a more accurate imaging decoding. First, the log-likelihood-ratios (LLRs) for $y_i$ relative to all possible integers values is calculated as\citep{davey1998low},
\begin{equation}\label{soft_LLR}
      L(y_i)=ln\frac{P(y_i=m\overline{y_0})}{P(y_i=0)}=\frac{(2y_i-m\overline{y_0})m\overline{y_0}}{2\sigma^2}
\end{equation} The maximum LLR $L_m$ calculated by equation (\ref{soft_LLR}) will indicate which integer $m_i$ is the most likely integer estimate for $y_i$. Then the likelihood for $y_i$ in GF(2) is obtained as,
\begin{equation}\label{soft_LLR2}
      L^{GF(2)}_i=
      \begin{cases}
        + \Delta L, &\text{when $m_i$ is odd}\\
        - \Delta L, &\text{when $m_i$ is even}\\
      \end{cases}
\end{equation}and,
\begin{equation}\label{soft_LLR2}
      \Delta L=L_m-\overline{L_m}
\end{equation}where $\overline{L_m}$ is the second largest LLR value calculated by equation (\ref{soft_LLR}).


The obtained $L^{GF(2)}_i$ values are used to decode the object information by applying decoding algorithms from communications. Here, assuming a block code such as low-density-parity-check (LDPC) code \citep{gallager1962low} or Luby Transform (LT) code \citep{luby2002lt}has been applied in the imaging encoding operation, the belief-propagation (BP) decoding algorithm\citep{mackay1995good,yue2014distributed} is given below as an example to explain the recovery procedure of the object information.

Taking advantage of the relationships expressed by solid lines in the bipartite graph shown in Fig. \ref{BP_explaination_Code} and using the LLRs of the received signal values obtained from remapping as \emph{a priori} information, the basic idea behind the BP algorithm is to obtain estimates of the pixel transmissivity by iteratively exchanging messages between illuminated pixel nodes and signal nodes\citep{pearl1982reverend}. First, $M_{j,i}$ representing the message passing from the $i$th signal node to the $j$th pixel node is initialized by,
\begin{equation}\label{LLR_initialization}
  M_{j,i}=L^{GF(2)}_i
\end{equation} Then in each BP iteration, $E_{j,i}$ representing the message passing from the $j$th pixel node to $i$th signal node is firstly calculated by using other indirect connected signal nodes,
\begin{equation}\label{BP_E}
  E_{j,i}=2tanh^{-1}(\prod_{i{'}\in B_j,i{'}\neq i}tanh(M_{j,i{'}}/2))
\end{equation} where $B_j$ is the set of the received signal nodes connected to the $j$th pixel node and,
\begin{equation}\label{tanh}
  2tanh^{-1}(p)=log\frac{1+p}{1-p}
\end{equation} Then $M_{j,i}$ is updated by summing up messages from all the indirect connected pixel nodes and its initial value,
\begin{equation}\label{BP_object_LLR}
    M_{j,i}=L^{GF(2)}_i+\sum_{j^{'} \in A_i,j^{'}\neq j}E_{j^{'},i}
\end{equation} where $A_i$ is the set of the illuminated pixel nodes that connected to the $i$th signal node. By iteratively repeating the calculation of $M_{j,i}$ and $E_{j,i}$, messages are generated and exchanged until the stopping criterion has been reached, i.e. either when a correct sequence is decoded or a maximum number of iterations has been reached. After the BP algorithm has been terminated, the resulting LLR of the $j$th pixel node is expressed as,
\begin{equation}\label{BP_result_LLR}
    L_j=\sum_{i\in B_j}M_{j,i}
\end{equation} and the resulting LLR of the $i$th received signal node is expressed as,
\begin{equation}\label{BP_result_LLR2}
    S_i=L_i^{GF^{(2)}}+\sum_{j\in A_i}E_{j,i}
\end{equation} Then the final decoded value $I_j$ for the $j$th pixel is determined by,
\begin{equation}\label{Final_result}
      I_j=
      \begin{cases}
        1,& \textmd{if}~L_j\geq0\\
        0,& \textmd{if}~L_j<0\\
      \end{cases}
\end{equation}

Apparently that the obtained a posteriori probability $L_j$ for the $j$th pixel value is determined not only based on the messages $L^{GF(2)}_i$ provided by directly connected received signal nodes, but also extra messages $\sum_{j^{'} \in A_i,j^{'}\neq j}E_{j^{'},i}$ obtained from other indirectly connected ones. For example, the information of the 4th pixel shown in Fig. \ref{BP_explaination_Code} is directly combined into the 1st and 2nd received signal values $y_1$ and $y_2$. Since $y_1$ also contains information from the 7th pixel, $y_2$ also includes information from 6th and 12th pixels, thus the information of the 4th pixel can be recovered with both the \emph{a priori} information provided by $y_1$ and $y_2$, as well as messages passing from other received signal values corresponding to the 6th, 7th and 12th pixels, namely $y_4$, $y_7$ and $y_8$. Consequently, even though $y_1$ and $y_2$ may be corrupted by noise, given a proper design of the ECC code, the information of the 4th pixel can still be recovered with a high probability. In other words, with the assistance from ECC codes, the object information can be protected from errors induced by noise.

When the decoding of the current received encoded sequence is finished, the decoder sends a request to the ECC encoder for sending the next encoded sequence when rateless codes are applied, while for fixed rate codes, the decoder resumes the decoding of the next received encoded sequence.

\section*{A Proof-of-concept Experiment}\label{experiment}
\begin{figure}
  \centering
  \includegraphics[width=5in]{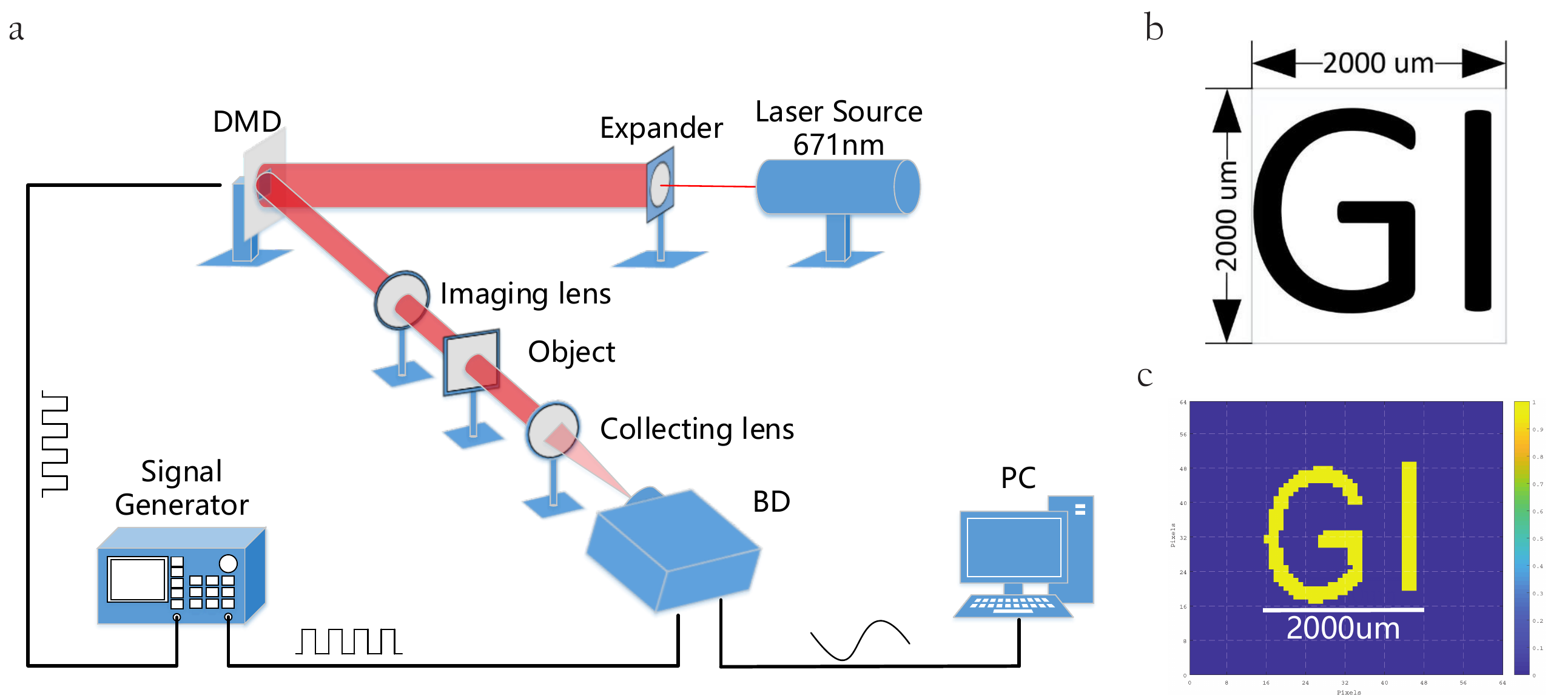}
  \caption{\textbf{Experiment setup and initial reconstruction result of target object.} \textbf{a}, A DMD is used to generate light fields with ECC structured patterns to encode the object information. The resulting encoded products are collected by the BD afterwards. \textbf{b}, The target object contains a \emph{GI} shaped hollow structure. \textbf{c} The reconstruction result of the object under SNR=2dB.}\label{TargetAndResult}
\end{figure}

A proof-of-concept experiment is employed to validate the effectiveness of our approach. As shown by the experiment setup in Fig. \ref{TargetAndResult}a, the testbed contains a continuous laser source, a DMD, an object, a charge-coupled-device (CCD) and a signal generator. The CCD here is used as a BD while the signal generator is used to provide synchronization between the DMD and BD. The laser source first generates a beam with a wavelength of 671nm. The beam is further adjusted by a beam expander and then uniformly projected on the DMD. By modulating the \emph{on-off} status of micromirrors on the DMD, a series of light fields are generated according to specific ECC encoding rules and projected on the object to perform encoding. With an additional collecting lens, the resulting light signals are collected by the BD afterwards and sent to the PC for decoding.

In order to satisfy the binary-valued quantization model, we employed an object with a hollow structure in our experiment. As shown in Fig. \ref{TargetAndResult}b, it contains a \emph{GI} shaped hollow structure with a size of $2000\mu m\times 2000\mu m$. Only the light that falls into the hollow part can pass through and reach the BD. We divide the target imaging plane where object is located into $64\times 64$ pixels and assume the pixel values corresponding to the hollow part are 1. Then LT code is employed in our experiment to encode the whole $K=4096$ pixels. To perform an LT-based encoding procedure, we first choose a number $d$ from a pre-designed probability distribution $\Omega(d)$, and then randomly select $d$ pixels to illuminate. After we performed $N=8192$ times of illuminations, an image of the object is obtained by using the aforementioned BP decoding algorithm. As we can see from Fig. \ref{TargetAndResult}c, the image of object is successfully reconstructed.
\begin{figure}
  \centering
  \includegraphics[width=3.5in]{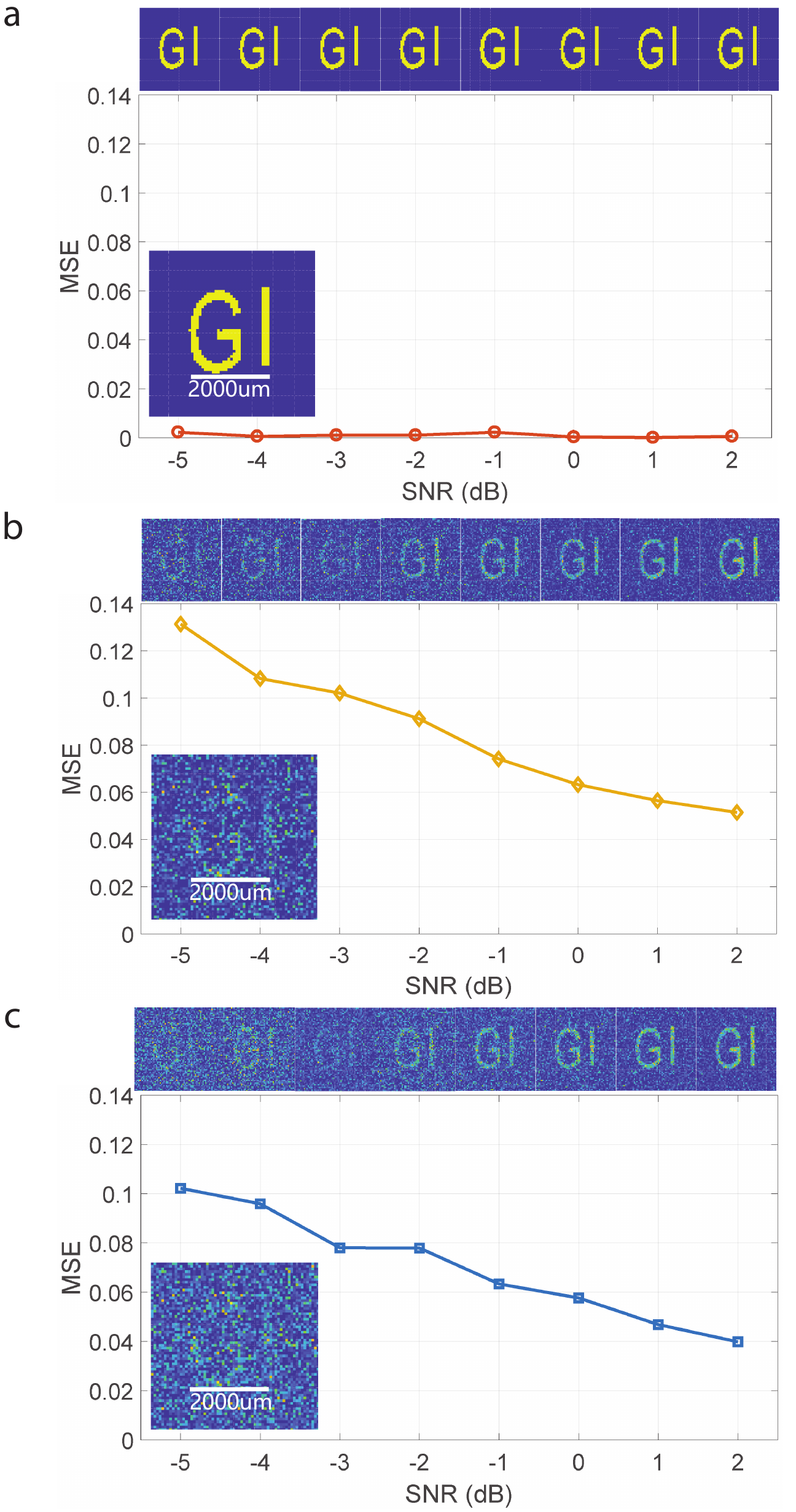}
  \caption{\textbf{Reconstruction images and MSE curves with SNR ranging from 2dB to -5dB.} \textbf{a} The reconstruction result by using the demonstrated approach. The object is stably reconstructed, indicating the object information is protected against the noise with the assistance from LT code. The reconstruction result when SNR=-5dB is presented in the insets. \textbf{b} and \textbf{c}, The reconstruction results by using conventional GI method. The object is heavily corrupted by the noise and becomes unrecognizable when SNR is lower than -3dB. The reconstruction results when SNR=-5dB are presented in the insets.}\label{MSE_LT_C_CS}
\end{figure}
As a further validation of the image quality improvement offered by the LT code, we gradually reduced the received signal-to-noise-ratio (SNR) in the reception by adding optical attenuators. First we define the SNR in our experiment as,
\begin{equation}\label{SNR_definition}
  \textmd{SNR}=10\textmd{log}_{10}\frac{P_s-P_n}{P_n}
\end{equation} where $P_s$ is the average power of the received signals and $P_n$ is the average power of background noise in the BD. Then we employ the mean-square-error (MSE) as a quantitative evaluation metric here for the reconstruction quality assessment \citep{lehmann2006theory},
\begin{equation}\label{MSEcalculation}
  \textmd{MSE}=\frac{1}{K}\sum_{j=1}^{K}(I_j-\overline{I_j})^2
\end{equation} where $K=4096$ is the total number of pixels under the current $64\times64$ resolution setting, $I_j$ is the $j$th pixel value on the obtained image and $\overline{I_j}$ is its original value. The obtained reconstruction results of the target object are presented in Fig. \ref{MSE_LT_C_CS}a. It can be seen that images of the target object are reliably reconstructed with the SNR ranging from 2dB to -5dB. Although some distortions can be observed when the SNR is down to -5dB, the major structure of the target object is still successfully reconstructed, indicating the object information is effectively protected against receiver noise by an LT code. This conclusion is also confirmed by the obtained MSE curve, which shows that the reconstruction error remains close to 0 within the evaluated SNR range.




As a comparison, with the same system setup, SNR range and the total number of illuminations, GI is chosen here as a representative of a conventional \emph{uncoded} imaging approach. 8192 light fields with Bernoulli-distributed patterns \citep{watts2014terahertz} are generated by the DMD and used to illuminate the object. Then the received signal collected by the BD is processed with mutual-correlation \citep{zhang2009improving} and gradient-projection (GP) \citep{zhao2012ghost,gong2016three} based algorithms respectively to reconstruct the image. As we can see from Fig. \ref{MSE_LT_C_CS}b and c, images of the target object can also be recovered by using conventional GI schemes. However, the obtained images from both of the reconstruction methods contain significant distortions. When the SNR condition is down to -5dB, the target object has been completely corrupted by noise. A similar conclusion can also be drawn by the obtained MSE curves. Results from both of the two reconstruction algorithms possess a significantly larger initial MSE, compared to our demonstrated approach at $\textmd{SNR}=2\textmd{dB}$. As the SNR worsens, their MSE curves keep increase dramatically, indicating that the object information is heavily corrupted by the receiver noise.

\section*{Discussion}\label{discussion}
Compared to results obtained by GI, images reconstructed by our demonstrated approach indicate a significant quality improvement under the same system settings. Although the employment of Bernoulli-distributed structured fields is physically similar to the encoding procedure in our design, it is only served for the purpose of linearly relating the GI incoherent detection fields to received signals, which is still within the framework of conventional \emph{uncoded} imaging methods. In addition, the demonstrated approach is also different from existing noise reduction techniques, since the object information is directly protected during the physical imaging process in our design rather than compensating errors in post-processing.

As seen from the experimental setup in Fig. \ref{TargetAndResult}a, the demonstrated approach possesses the same implementation complexity as a typical GI imagery. However, its reconstruction complexity may vary as the applied encoding and decoding scheme changes. For example, when a \emph{hard-decision} based non-iterative decoding method is used, the complexity of the demonstrated approach will be around $O(KlnK)$\citep{mackay2005fountain}, which is lower than the mutual-correlation based GI algorithm whose complexity is around $O(K^3)$. But when an iterative decoding method is applied, the convergence of our demonstrated approach will be determined by a variety of factors. The total number of illuminations, SNR conditions and pre-defined maximum iteration limits will all affect the termination of the BP decoding algorithm. Besides, the convergence of the BP algorithm is also related to the number of pixels involved in one encoding operation. Similarly, the termination of a GP based iterative algorithm in GI depends on the number of measurements, SNR conditions and pre-defined maximum iteration limits. Moreover, its convergence is also affected by the initial error and the condition number of the projection matrix\citep{figueiredo2007gradient}.

Besides, several aspects still need to be further investigated in our future work. In the current approach, transmissivity is selected to represent the object information and a binary-valued model is assumed. Other representations such as reflectance and refraction\citep{barrett2013foundations}, as well as quantization models such as grey-scaled ones can also be applied here. For illustration purpose, the AWGN noise from components in BD is assumed in our experiment. Thus other error sources such as Poisson-distributed photon noise and Rayleigh-distributed fading\citep{barrett2013foundations}, as well as atmosphere turbulence\citep{zhang2010correlated} need to be further investigated. In addition, ECC codes are originally designed and optimized for communication channels. In the next step, optimizations of the mathematical structure of ECC codes in the application of imaging are to be conducted.

\section*{Conclusion} \label{conclusion}
In this article, we demonstrate the application of ECC codes used in communication systems for improving the quality of an imaging system. In our design, the ECC encoding process is physically realized by using light fields generated by a DMD while the decoding of the object information is performed upon the signal collected by a BD. Our proof-of-concept experiment validates that ECC codes can provides effective protection of object information against receiver noise. The superiority of the demonstrated approach is further confirmed by benchmarking with conventional GI schemes under the same system setting and SNR conditions. We believe that the demonstrated approach opens up a new framework for developing a noise-resistive and highly-reliable imaging technology: \emph{ECC Assisted Imaging}. This technology could bring fundamental paradigm shifts in many applications such as remote sensing, spectroscopy and biomedical imaging.

\section*{Methods}
\noindent\textbf{Experimental details.} In the present proof-of-concept experiment, the source was a 671nm semi-conductor continuous laser source. A beam expander was applied to uniform the laser beam. A Texas Instruments V-7001 DMD with $1024\times768$ micromirrors in total was used in the experiment, but only $256 \times 256$ located at the center were employed to generate light fields. The size of each micromirror was $13.6\mu m\times 13.6\mu m$. Thus the overall size of generated light fields was $3481\mu m\times3481\mu m$. The generated light fields were projected on the hollow structured object by using a 1:1 imaging lens. Then the transmit-through lights were collected by an AVT Stingray F-504C CCD with the assistance from a collecting lens. For synchronization between the DMD and CCD, a Tektronix AFG3252C Arbitrary Function Generator was used to generate a square wave with a frequency of 20MHz. For the control of SNR conditions, we used 25mm Absorptive ND Filter Kit from Thorlabs.

\noindent\textbf{Distribution $\Omega(d)$.} The distribution $\Omega(d)$ that employed in the encoding procedure in our proof-of-concept experiment is defined as follows,\\
(1) The range of $d$ is from 1 to 10;\\
(2) The probability of $d=1$ and $d=2$ is $30\%$ respectively;\\
(3) The probability for each of the rest values is $5\%$.

\bibliographystyle{unsrt}
\newpage
\bibliography{Journal_main}
\newpage
\section*{Author contributions}
X.W. conceived the original idea and designed the LT code for experiment; Z.B. designed and assembled the optical testbed; X.W. and Z.B. conducted the experiments and analyzed the data; Z.L. and W.G. supervised the work; Z.L., W.G., B.V. and S.H. conducted the discussions of the results; X.W. and B.V. wrote the manuscript and all authors reviewed the manuscript. 

\end{document}